\documentclass{ws-mpla}

\newcommand{\Slash}[1]{\ooalign%
{\hfil\rotatebox{30}{\underline{\hspace*{0.6cm}}}\hfil\crcr$#1$}}

\begin{document}

\markboth{Masayasu Harada and Chihiro Sasaki}
{Dilepton Production from Dropping $\rho$ in the VM}

%%%%%%%%%%%%%%%%%%%%% Publisher's Area please ignore %%%%%%%%%%%%%%
\catchline{}{}{}{}{}
%%%%%%%%%%%%%%%%%%%%%%%%%%%%%%%%%%%%%%%%%%%%%%%%%%%%%%%%%%%%%%%%%%%

\title{Dilepton Production from Dropping $\rho$
in the Vector Manifestation\footnote{%
Talk given by M.~Harada at 
``Chiral Symmetry in Hadron and Nuclear Physics (Chiral 07)''
(November 13-16, 2007, Osaka University, Japan).
This is based on the work done in Ref.~[1].
}
}

\author{\footnotesize Masayasu Harada}

\address{Department of Physiccs, Nagoya University
Nagoya, 464-8602, Japan}

\author{Chihiro Sasaki}

\address{
Physik-Department, Technische
Universit\"at M\"unchen, D-85747 Garching, Germany
}

\maketitle

%\pub{Received (Day Month Year)}{Revised (Day Month Year)}

\begin{abstract}
In this write-up we summarize the main result of 
our analysis on the thermal dilepton
production rate from the
dropping $\rho$ based on the vector manifestation (VM).
In the analysis, we showed that the
effect of the strong violation of the vector dominance (VD)
predicted by the VM,
substantially suppresses the dilepton production rate
compared with the one predicted by assuming the VD together 
with the dropping $\rho$.
\end{abstract}

\ccode{PACS Nos.: 12.39.Fe, 12.40.Vv}

\vspace{0.2cm}

\nocite{HS:HLSDL}

Changes of hadron properties are indications of chiral symmetry 
restoration occurring in hot and/or dense QCD and have been 
explored using various effective chiral approaches~\cite{rest,RW}.
An enhancement of dielectron mass spectra below
the $\rho / \omega$ resonance was first observed at CERN 
SPS~\cite{ceres}, which can be
explained by the
dropping masses of hadrons following the Brown-Rho (BR)
scaling~\cite{BR-scaling} (see e.g., Ref.~\refcite{RW}).

The vector manifestation (VM)~\cite{HY:VM} 
is a novel pattern 
of the Wigner realization of chiral symmetry in which
the $\rho$ meson becomes massless degenerate with the
pion at the chiral phase transition point.
The VM is formulated~\cite{HY:PRep,HS:VM,HKR:VM,Sasaki:D} 
in the effective field theory based
on the hidden local symmetry (HLS)~\cite{BKY},
and thus
gives
a field theoretical description of the dropping $\rho$ mass.

The dropping mass is supported by
the mass shift of the $\omega$ meson in nuclei measured by
the KEK-PS E325 Experiment~\cite{KEK-PS} and
the CBELSA/TAPS Collaboration~\cite{trnka}.
Furthremore,
recent Phenix data cannot be explained by a hadronic 
model~\cite{Phenix}, which might indicate changes of some properties 
of vector mesons.
It
seems difficult to explain the dimuon data from NA60
by a naive dropping $\rho$~\cite{NA60}.
However, the strong violation of the vector dominance (VD)
is not considered,
which is one of the significant predictions of the VM~\cite{HS:VD}
and plays an important role~\cite{Brown:2005ka-kb}.

In Ref.~\refcite{HS:HLSDL},
we studied the dilepton production rate
from the dropping $\rho$ based on the VM
using the HLS theory at finite temperature.
We paid a special attention to 
the effect of the violation of the vector dominance
(indicated by ``$\Slash{\rm VD}$'') 
which is due to the
{\it intrinsic temperature effects} of the parameters
introduced through the matching to QCD in the Wilsonian
sense.
We made a comparison of the dilepton production rates
predicted by the VM with the ones by the dropping $\rho$
under the assumption of the VD.
The result shows that the effect of the $\Slash{\rm VD}$
substantially suppresses the dilepton production rate
compared with the one predicted by assuming the VD together 
with the dropping $\rho$.

The VM was
formulated~\cite{HY:PRep,HS:VM,HKR:VM,Sasaki:D} 
in the effective field theory based
on the HLS~\cite{BKY}.
At the leading order 
the HLS Lagrangian includes three parameters:
the pion decay constant $F_\pi$; the HLS gauge coupling $g$;
and a parameter $a$.
Using these three parameters, the $\rho$ meson mass $m_\rho$
and
the direct $\gamma$-$\pi$-$\pi$ coupling strength $g_{\gamma\pi\pi}$
are expressed as
$m_\rho^2 = g^2 a F_\pi^2$
and 
$g_{\gamma\pi\pi} = 1 - \frac{a}{2}$.
{}From these expressions, one can easily see that the 
VD of the electromagnetic form factor of the pion,
i.e. $g_{\gamma\pi\pi}=0$, is satisfied for $a = 2$.

The most important ingredient to formulate the VM in hot matter
is the following intrinsic temperature dependences of 
the bare parameters $a$ and $g$~\cite{HS:VM,Sasaki:D}:
\begin{equation}
g(\Lambda;T) \sim \langle \bar{q}q \rangle \rightarrow 0 \ ,
\quad
a(\Lambda;T) - 1 \sim \langle \bar{q}q \rangle^2 \rightarrow 0 \ ,
\label{bare g a}
\end{equation}
for $T \rightarrow T_c$.
As a result, the vector meson pole mass also goes to zero
for $T \to T_c$:
\begin{equation}
m_\rho(T) \sim \langle \bar{q}q \rangle \to 0\,.
\end{equation}
We would like to stress that the VD
is strongly violated near
the critical point associated with the dropping $\rho$
in the VM in hot matter~\cite{HS:VD}:
\begin{equation}
a(T) \rightarrow 1 \ , 
\quad
\mbox{for} \ T \rightarrow T_c \ .
\end{equation}

We should note that
the conditions in Eq.~(\ref{bare g a}) hold
{\it only in the vicinity of $T_c$}:
They are not valid any more far away from
$T_c$ where ordinary hadronic temperature
corrections are dominant.
For expressing a temperature above which the intrinsic
effect becomes important,
we introduce a temperature $T_f$, 
so-called flash temperature~\cite{BLR}.
The VM and therefore the dropping $\rho$ mass become 
transparent for $T>T_f$.
On the other hand, we expect that
the intrinsic effects are negligible in the low-temperature
region below $T_f$:
Only hadronic temperature corrections are considered for $T < T_f$.
Based on the above consideration, we adopt the following
ansatz of the temperature dependences of the 
bare $g$ and $a$:
\begin{eqnarray}
  g(\Lambda;T) \propto \langle \bar{q}q \rangle_{T} 
\ , \quad
  a(\Lambda;T) - 1 \propto \langle \bar{q}q \rangle_{T}^2
\quad
\ \mbox{for}\ T > T_f \ ,
\label{bare t dep}
\end{eqnarray}
while $g(\Lambda;T)$ and 
$a(\Lambda;T)$ are constants for $T < T_f$.

\begin{figure*}
\begin{center}
\includegraphics[width = 3.8cm]{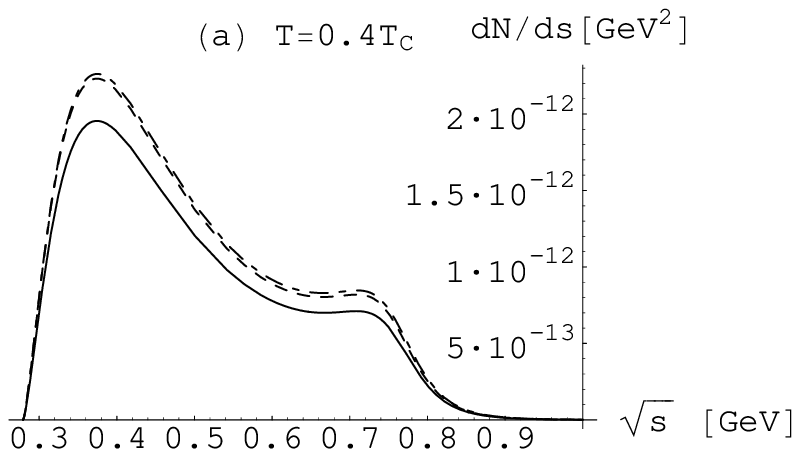}
\hspace*{0.5cm}
\includegraphics[width = 3.8cm]{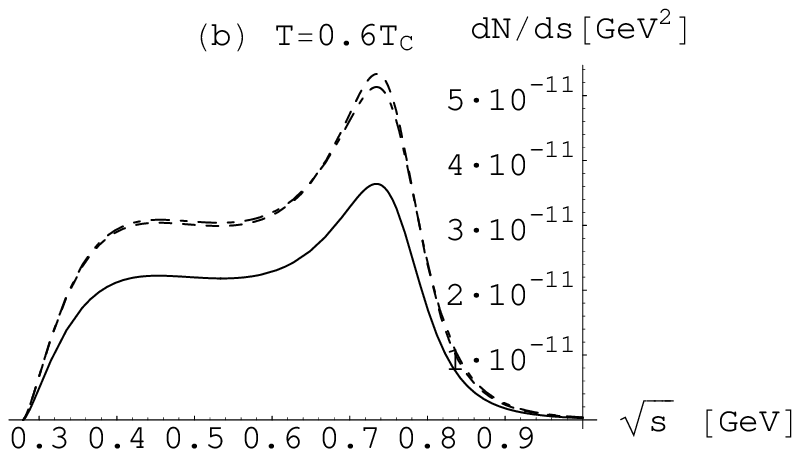}
\\
\includegraphics[width = 3.8cm]{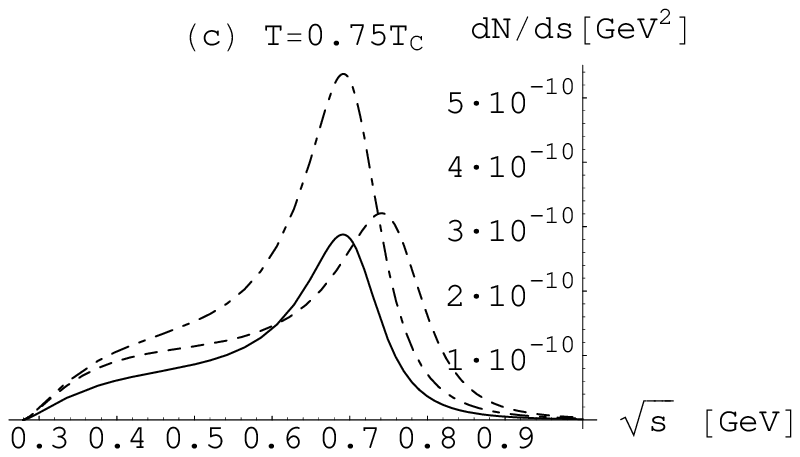}
\hspace*{0.2cm}
\includegraphics[width = 3.8cm]{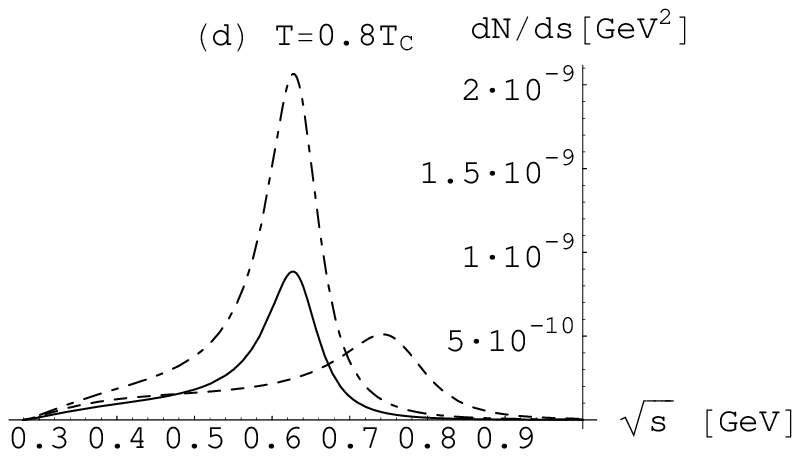}
\hspace*{0.2cm}
\includegraphics[width = 3.8cm]{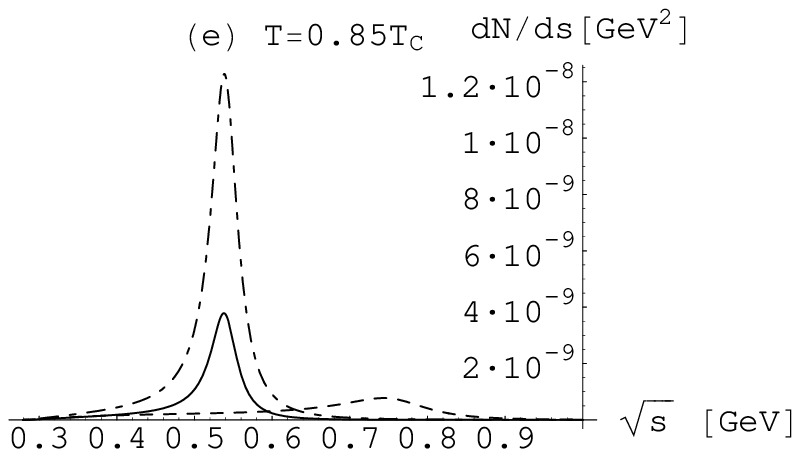}
\end{center}
\caption[]{
Dilepton production rate as a function of the invariant 
mass $\sqrt{s}$ for various temperatures.
The solid curves include the effects of the violation of the VD.
The dashed-dotted curves correspond to the analysis assuming
the VD. The dashed curves represent the result with 
the parameters at zero temperature.
}
\label{fig:dl}
\end{figure*}

As noted,
the vector dominance (VD)
is controlled by the parameter $a$ in the HLS theory.
The VM leads to the strong violation of the VD 
(indicated by ``$\Slash{\rm VD}$'') 
near the chiral symmetry restoration point, which can be traced 
through the Wilsonian matching and the RG evolutions.
Thus the direct photon-$\pi$-$\pi$ coupling $g_{\gamma\pi\pi}$
yields 
non-vanishing contribution to the form factor together with the 
$\rho$-meson exchange.
In Ref.~\refcite{HS:HLSDL},
we compared the dilepton spectra predicted in the VM
(including the effect of $\Slash{\rm VD}$) with those
obtained by assuming the VD, i.e. taking 
$g_{\gamma\pi\pi}=0$.
Figure~\ref{fig:dl} shows the form factor and the dilepton 
production rate integrated over three-momentum,
in which the results with VD and $\Slash{\rm VD}$
were compared.
%%%%%%%%%%%%%%%%%%%%%%%%%%%%%%%%%%%%%%%%%%%
%%%%%%%%%%%%%%%%%%%%%%%%%%%%%%%%%%%%%%%%%%%%%%%
The figure shows a clear difference between the curves
with VD and $\Slash{\rm VD}$.
It can be easily seen that the $\Slash{\rm VD}$ gives a reduction
compared to the case with keeping the VD.
The features of the dilepton production 
rate coming from two-pion annihilation shown in 
Fig.~\ref{fig:dl}(a)-(e) are summarized below for each temperature:
%%%%%%%%%%%%%%%%%%%%%%%%%%%%%%%%%%%

\noindent
{\bf (a) and (b) (below $T_f$) :}
\quad
In both (a) and (b),
the dilepton rates for $\Slash{\rm VD}$ (indicated by solid curves)
are suppressed compared with those for VD
(indicated by dashed-dotted curves).
This is due to decreasing of the $\rho$-$\gamma$ mixing strength
$g_\rho$ at finite temperature for $\Slash{\rm VD}$.
In case with VD, however, $g_\rho$ is almost constant, and the
dashed-dotted curves almost coicide with the dashed ones 
for the vacuum $\rho$.

\noindent
{\bf (c), (d) and (e) (above $T_f$) :}
\quad
A shift of
the $\rho$ meson mass to lower-mass region can be seen.
Furthermore, 
the production rate based on the VM 
(i.e., the case with $\Slash{\rm VD}$) is suppressed compared
to that with the VD.
We observe that the suppression is more transparent 
for larger temperature:
The suppression factor is $\sim 1.8$ in (c), $\sim 2$ in (d)
and $\sim 3.3$ in (e). 

As one can see in (c), the peak value of the rate
predicted by the VM for $T \gtrsim T_f$
is even smaller than the one obtained by the vacuum parameters,
and the shapes of them are quite similar to each other.
This indicates that it might be difficult to measure the 
signal of the dropping $\rho$ experimentally, if this
temperature region is dominant in the evolution.
In the case shown in (d), on the other hand,
the rate by the VM 
is enhanced by a factor of 
about two compared with the one by the vacuum $\rho$.
The enhancement becomes prominent near the critical temperature
as seen in (e).
These imply that we may have a chance to discriminate the
dropping $\rho$ from the vacuum $\rho$.

We cannot make a direct comparison of our results
with experimental data.
However a {\it naive} dropping $m_\rho$ formula, i.e., $T_f = 0$, 
as well as VD in hot/dense matter are sometimes used for
theoretical implications of the data.
As we have shown,
the violation of the VD gives a clear difference
from the results without including the effect.
It may be then expected that a field theoretical analysis
of the dropping $\rho$ as presented in this work and a reliable
comparison with dilepton measurements will provide an evidence
for the in-medium hadronic properties associated with the chiral
symmetry restoration, if complicated hadronization processes do
not wash out those changes.

Recently the chiral perturbation theory with including vector
and axial-vector mesons as well as pions has been 
constructed~\cite{HS:GHLS}.
It is interesting to see the effect of inclusion of the
axial-vector meson to the dilepton rate~\cite{HS:DL2}.

\vspace{0.2cm}

\noindent
{\bf Acknowledgments}

We are grateful to Gerry Brown, Bengt Friman and Mannque Rho 
for fruitful discussions and comments.
We also thank Jochen Wambach for stimulating discussions.
The work of C.S. was supported in part by the Virtual Institute
of the Helmholtz Association under the grant No. VH-VI-041.

\end{document}